%% file: main.tex
\documentclass{article}
\usepackage{spconf,amsmath,graphicx}

\usepackage{subcaption}
\usepackage{booktabs}
\usepackage{array}
\usepackage{enumitem}
\usepackage{multirow}
\usepackage{url}

\newcommand{\ours}{RedPen}

\title{RedPen: Region- and Reason-Annotated Dataset of Unnatural Speech}
\name{Kyumin Park$^{1}$, \quad Keon Lee$^{12}$\sthanks{This work was done when Keon Lee was with KAIST as a MS student.}, \quad Daeyoung Kim$^{1}$, \quad Dongyeop Kang$^{3}$}
\address{
$^{1}$School of Computing, KAIST, Republic of Korea \\
$^{2}$KRAFTON Inc., Republic of Korea \\
$^{3}$University of Minnesota, US \\
\texttt{\{pkm9403, kimd\}@kaist.ac.kr, keonlee@krafton.com, dongyeop@umn.edu}
}

\begin{document}
%
\maketitle
\input{content/00_abstract}
\input{content/01_introduction}

\input{content/02_data_collection}
\input{content/04_analysis}
\input{content/06_conclusion}
\bibliographystyle{IEEEbib}
\bibliography{anthology,custom}

\end{document}

%% file: content/00_abstract.tex
\begin{abstract}
Even with recent advances in speech synthesis models, the evaluation of such models is based purely on human judgement as a single naturalness score, such as the Mean Opinion Score (MOS).
The score-based metric does not give any further information about which parts of speech are unnatural or why human judges believe they are unnatural.
We present a novel speech dataset, RedPen, with human annotations on unnatural speech \textit{regions} and their corresponding \textit{reasons}.
RedPen consists of 180 synthesized speeches with unnatural regions annotated by crowd workers; These regions are then reasoned and categorized by error types, such as voice trembling and background noise.
We find that our dataset shows a better explanation for unnatural speech regions than the model-driven unnaturalness prediction.
Our analysis also shows that each model includes different types of error types.
Summing up, our dataset successfully shows the possibility that various error regions and types lie under the single naturalness score. We believe that our dataset will shed light on the evaluation and development of more interpretable speech models in the future.
Our dataset will be publicly available upon acceptance.

\end{abstract}

%% file: content/01_introduction.tex
\section{Introduction}
\label{sec:introduction}

Naturalness is one of the most influential factors in evaluating speech models \cite{Wang2017tacotron} on whether or not output speech naturally sounds like a human. 
The naturalness of speech models is often quantified by human evaluation with single-score metrics such as Mean Opinion Score (MOS), or the human judgement is estimated by deep learning-based predictor \cite{Lo2019, mittag2019non, Mittag2020}.

When scoring naturalness, human judges listen to each audio and give a single score to the audio. 
However, a single, unified score does not provide further information but only the overall naturalness of the entire speech. It does not provide where and why the score is determined. 
In fact, there exist several factors that affect human perception of naturalness, such as speech style \cite{dall2014rating} or acoustic features including pitch and energy \cite{omahony21_ssw, Baird2018perception}.

To better interpret the model's behavior and find such factors contributing to the final naturalness, various interpretation methods on deep learning models have been studied \cite{simonyan2014deep, sundararajan2017axiomatic}.
However, a recent study shows that these saliency-based measurements cannot fully represent human perception in detecting linguistic styles \cite{hayati-etal-2021-bert}.
This motivates us to collect human's real perceptions for speech naturalness, and propose them as interpretable measurements for speech evaluation.

\input{supplementary/redpen_main}

In this paper, we introduce \ours{}, a region- and reason-annotated dataset of synthesized speeches. 
We developed a tool asking people to annotate unnatural regions in each audio. From the annotations, we additionally categorize each annotated region into common error types in speech synthesis, as shown in Figure \ref{fig:redpen_main}.
Our analyses find that our region annotation represents unnatural regions better than the predicted interpretation from the MOS prediction model.
Also, the reason for unnaturalness varies by different speech synthesis models.
In our human evaluation, our dataset is judged better than the previous system in providing better explanations of naturalness. Through our dataset, we show evidence of several factors lying under a single naturalness score. 

\input{supplementary/sample_5}

Our Contributions are as:
\begin{itemize}[noitemsep,topsep=0pt]
    \item suggest a novel dataset containing unnatural regions with reasons among 11 types on 180 synthesized speeches.
    \item show that model interpretation is different from human perception on unnatural regions. 
\end{itemize}

%% file: supplementary/redpen_main.tex
\begin{figure}[t]
    \centering
    \includegraphics[width=.45\textwidth]{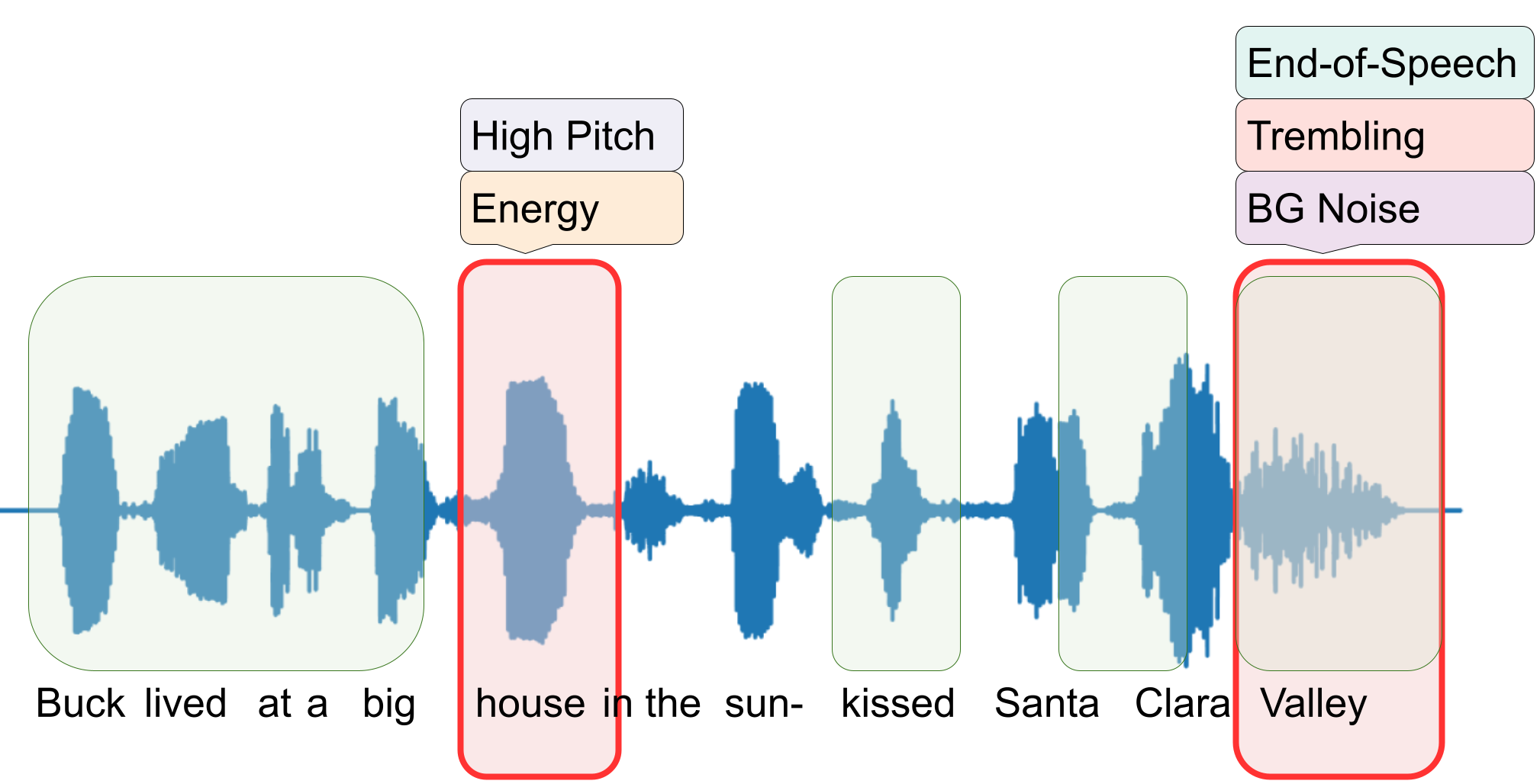}
    \caption{The sample annotation of unnatural regions. There is a discrepancy between the human-annotated unnatural regions (red) with their reason categories and salient regions predicted from the model (green).}
    \label{fig:redpen_main}
\end{figure}

%% file: supplementary/sample_5.tex
\begin{figure*}[t]
    \centering
    \includegraphics[width=.9\textwidth]{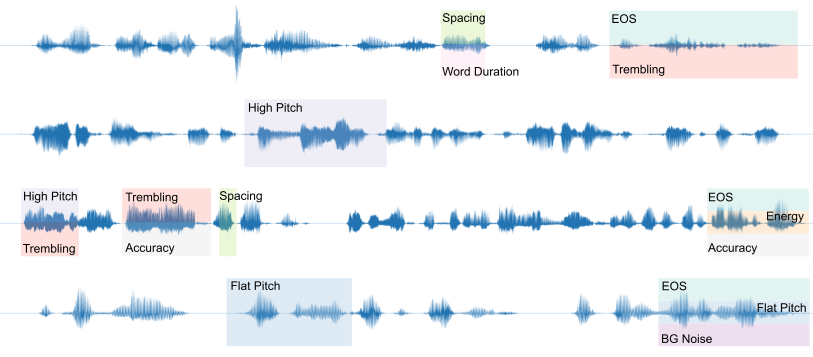}
    \caption{The sample region- and reason-annotations. Each color indicates distinct error type labels. Each speech span could be annotated with multiple reasons (e.g., EOS, BG Noise, and Flat Pitch for the last part of the last speech above)}
    \label{fig:redpen_sample_5}
\end{figure*}

%% file: content/02_data_collection.tex
\section{RedPen: Annotated Dataset on Unnatural Speeches}
\label{sec:data_collection}

Table \ref{tab:basic_stats} shows data statistics of our dataset. 
We annotate the total 180 audio samples. 
We sample synthesized speeches from the VCC2018 challenge \cite{Lorenzo-Trueba2018vcc}, which consists of synthesized speeches from various voice conversion systems and their evaluation scores.
Figure \ref{fig:redpen_sample_5} shows some example annotations in our dataset.
We describe the region annotation (\S\ref{subsec:region}) and reason annotation (\S\ref{subsec:reason}) in the following sections. Our dataset will be publicly available.

\subsection{Region Annotation}\label{subsec:region}
We first collect information about unnatural regions from crowd workers.
Using our annotation tool, we ask workers to annotate unnatural region in given audio. Each worker annotates 10 audios, including one human-recorded audio for filtering bad annotation.
We used Amazon Mechanical Turk\footnote{\url{https://mturk.com}} to recruit participants and invite them to do annotation tasks in our annotation tool. 

Human annotations on speeches are a set of continuous regions, so they are hard to align with discrete labels.
Thus, we use a binning-based calculation of inter-annotator agreement; more specifically, bin annotation divides each speech into 0.1s bin and marks each bin annotated when time annotation overlaps the bin. 
\ours{} consists of three participants' annotations for each speech. 

Inter-annotator agreement of our region annotation is 0.697, which is regarded to be reliable according to  \cite{landis1977measurement}. Also, the correlation between region length and overall score indicates that the longer unnatural regions are in audio, the more unnatural speech it is. Since the regions are annotations of unnatural segments, the negative correlation between region length and audio naturalness is reasonable.

\input{supplementary/basic_stats}



\textbf{Model-interpreted salient regions.}
As a baseline interpretation model to compare against our human annotations, we use the MOS prediction model and obtain their salient regions using Integrated Gradient \cite{sundararajan2017axiomatic}. 
The MOS prediction model is trained using a self-supervised learning model \cite{cooper2022generalization} trained on synthesized speeches from the VCC2018 challenge. 
We randomly split the speech corpus into training and test splits, but all human-annotated speeches are included only in the test split. 
Since saliency scores are calculated in a frame unit, these interpretation scores are max-pooled to 0.1s unit.

Figure \ref{fig:redpen_main} shows an example speech in our dataset, with human-annotated regions (red) and model-predicted salient regions (green).
In the analysis section, we will provide the correlation analysis of these two regions.

\subsection{Reason Annotation} \label{subsec:reason}

\input{supplementary/categories}

Annotated regions might be more explainable than a single score, but we still do not know the underlying reasoning behind why humans perceive these regions as unnatural. 
Therefore, we annotate the selected regions with their corresponding error types (also called reasons). 
We first categorize every single annotation into common error types in speech synthesis.

To determine the categories, we consider features known to correlate with naturalness. Extending \cite{Baird2018perception} showing that pitch variance correlates with naturalness, we set categories of acoustic feature errors: \textit{pitch}, \textit{energy}, and \textit{duration} because these features are used to determine speech style \cite{ren2019fastspeech, lee21styler}. 
Considering intelligibility as another objective measurement of speech synthesis \cite{ullmann2015objective}, we add \textit{accuracy} category, which denotes the wrong prediction of phonemes. We also add \textit{end-of-speech} category which related to error propagation problem in speech synthesis models \cite{ren2019fastspeech} and \textit{background noise} which should be considered to synthesize clean speech \cite{lee21styler}. The final categories are listed in Table~\ref{tab:categories} with their short descriptions.

Using the categories, we label each annotated region to each error category. We allow multi-labels since several error types can occur in a single erroneous region. Error type labeling is conducted by speech-expertises to avoid crowd workers' bias on error types.
The average number of error types per region annotation is 1.52 (See Table\ref{tab:basic_stats})


%% file: supplementary/basic_stats.tex

\begin{table}[t]
    \centering
    \small
    \begin{tabular}{cc}
    \toprule
        \# Audio samples & 180 \\
        Mean audio duration (s) & 3.444 \\
        \hline
        \# region annotations per sample & 1.459 \\
        Mean region annotation length (s) & 0.422 \\
        \# reason labels per annotation & 1.522 \\
        Inter-annotator agreement on regions & 0.697 \\
        $Corr(l, s)$ & -0.687 \\
    \bottomrule
    \end{tabular}
    \caption{Data statistics of RedPen. Inter-annotator agreement score is average score of Cohen's Kappa agreement using bin annotations. $Corr(l, s)$ is the correlation between overall score and length of unnatural regions. Negative correlation indicates longer unnatural regions are annotated in more unnatural speech.}
    \label{tab:basic_stats}
\end{table}

%% file: supplementary/categories.tex
\begin{table}[t]
    \centering\small
    \begin{tabular}{@{}rp{.30\textwidth}@{}}
    \toprule
        \textbf{Error Type} & \textbf{Description} \\
    \midrule
        End of Speech & Position of region is at the end of each utterance. \\
        
        Silence & Position of region is at the non-verbal region. \\
        
        High pitch & Pitch of region is higher than rest of audio. \\
        
        Voice Trembling & Voice is trembling in the region. \\
        
        Flat pitch & Pitch of erroneous region is significantly flat. \\
        
        Energy-related & Energy of the region is significantly larger or smaller than other regions. \\
        
        Spacing & Length of spacing is too short or too long. \\
        
        Word duration & Length of word / phoneme is too short or too long. \\
        
        Accuracy & Wrong phoneme is inserted or some phonemes are missed. \\
        
        Background noise & Noisy sound is inserted in background.  \\
        
        Undefined & None of the above reasons. \\
    \bottomrule
    \end{tabular}
    \caption{List of error types occurring audio unnaturalness and their explanations}
    \label{tab:categories}
\end{table}

%% file: content/04_analysis.tex
\section{Analysis on Annotations}
\label{sec:region_analysis}

To function as an interpretable evaluation metric, region annotation should represent unnatural regions accurately, and reason annotations should reflect the underlying error types of the annotated regions. 
To validate the effectiveness of RedPen, we perform the following analyses.

\input{supplementary/captum_human_comparison}
\input{supplementary/model_human_correlation}

\subsection{Comparison with Model Interpretation}
We compare human-annotated regions to model-predicted salient regions by the naturalness prediction to observe which method represents unnatural regions better. 



We conduct a preference test between human annotations and model interpretations. We use overlapping regions and the union of regions as human annotations. 
Table~\ref{tab:region_comparison} shows region comparison between human annotation and model interpretation. Union of human annotation is more preferred than model interpretation, while overlap is equally preferred as model interpretation. Overlap is less preferred than union because overlap becomes sparse easily. Indeed, overlap occupies 4\% of total audio while union possesses 40\%.

Next, we observe the overlap between model interpretation and human annotation to verify whether any relevance exists between the two. 
We compute Pearson's correlation between model interpretation and human annotations (bin annotation) for each error type and the overall region. 

Table~\ref{tab:model_human_corr} describes the correlation between model interpretation and annotated regions for each error type and overall human annotation. The correlation value is near 0 for all error types and overall, which indicates that model interpretation is irrelevant to any of the error types. This indicates that model interpretation cannot represent any error type.

\subsection{Unnaturalness Comparison between Models}

\input{supplementary/category_dist}

To measure the model's frequent error types, we aggregate annotated synthesized speeches by models. 
We only include the models containing at least 10 samples in our annotations, and all distributions are normalized by the number of samples in \ours{}. 

As depicted in Figure~\ref{fig:category_dist}, we find that \textit{System1} reports a large number of trembling errors, while it has a small number of duration and spacing errors compared to the average. On the other side, \textit{System2} reports more accuracy errors than the other models, while the average models mostly have end-of-speech and trembling errors predominantly. This indicates that error type varies by model. 
Within this result, we suggest that our annotation can be used as a detailed evaluation metric. With error type labeling added to our annotation, developers can induce the strength and weaknesses of the model.

\subsection{\ours{} for speech evaluation benchmark}
Imagine we scale up \ours{} on the current speech benchmarks or develop an automatic model to predict these regions and reasons.
Then, \ours{} could be used as a new benchmark for evaluating speech synthesis methods.
To check the possibility, we perform a human evaluation on different versions of \ours{} by asking human annotators to choose which version of the system they prefer to use in the speech evaluation.
We compare three systems representing from \ours{}: single score only, score \& region, score \& region \& reason. We sample 50 annotations randomly from \ours{}, and three people select the preferred system for each sample.


\input{supplementary/system_preference}

Table \ref{tab:system_prefer} shows the preference test result. People prefer a system with scores and regions the most, followed by a system with all annotations and a system with only a single number. Reason annotation is less preferred than region annotation only.
Even though reason annotation can provide helpful information to model developers, complicated evaluation procedures would be less preferred by people.

%% file: supplementary/captum_human_comparison.tex
\begin{table}[t]
    \centering
    \small
    \begin{tabular}{ll|c|c}
    \toprule
         & & \textbf{Human} & \textbf{Model} \\
    \midrule
        \multirow{2}{*}{Preference} & Union & 57.5\% & 42.5\% \\
         & Overlap & 49.5\% & 50.5\% \\
    \bottomrule
    \end{tabular}
    \caption{Preference test result comparing human annotation and model interpretation. Preference indicates people's preference on question \textit{'which region represents unnaturalness better?'}. }
    \label{tab:region_comparison}
\end{table}

%% file: supplementary/model_human_correlation.tex

\begin{table}[t]
    \centering
    \small
    \begin{tabular}{l|c|l|c}
    \toprule
        \textbf{Error Type} & \textbf{Corr.} & \textbf{Error Type} & \textbf{Corr.} \\
    \midrule
        EOS & 0.017 & Energy & 0.015 \\
        Duration & 0.009 & Spacing & 0.012 \\
        Silence & -0.011 & Accuracy & 0.019 \\
        High Pit. & 0.018 & BG Noise & 0.004 \\
        Trembling & 0.024 & Undefined & 0.005 \\
        Flat Pitch & 0.017 & \textbf{Overall} & 0.015 \\
    \bottomrule
    \end{tabular}
    \caption{Correlation coefficient between model interpretation score and region annotation for each error types.}
    \label{tab:model_human_corr}
\end{table}

%% file: supplementary/category_dist.tex

\begin{figure}[t]
    \centering
    \includegraphics[trim={0.5cm 0cm 0.5cm 0 },clip,width=.5\textwidth]{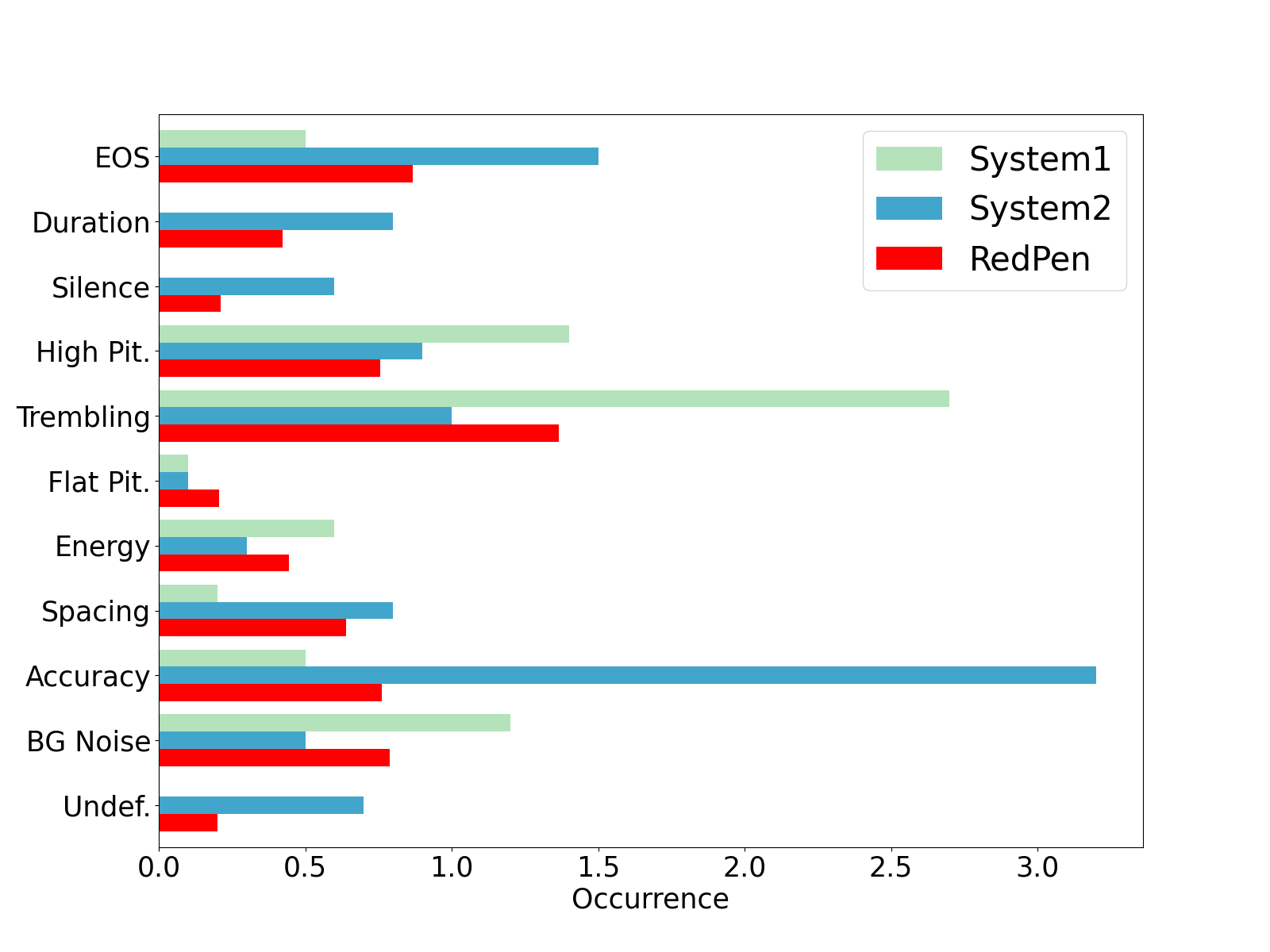}
    \caption{Error type distribution of entire dataset and of single model. y-axis denotes the category in Table~\ref{tab:categories}, and x-axis is the occurrence. The number of occurrence is normalized by dividing occurrence into the number of samples. System1 and System2 are submitted models in VCC2018 with code \textit{N18} and \textit{N05}, respectively. \ours{} is an average occurrence in all samples. Best viewed in colors.}
    \label{fig:category_dist}
\end{figure}

%% file: supplementary/system_preference.tex
\begin{table}[t]
    \centering
    \small
    \begin{tabular}{l|c}
    \toprule
        \textbf{System} & \textbf{Preference} \\
    \midrule
        Score & 26.7\% \\
        Score + Region & 41.3\% \\
        Score + Region + Reason & 32.0\% \\
    \bottomrule
    \end{tabular}
    \caption{Preference test on different versions of \ours{} on unnaturalness evaluation. Score is single naturalness score per audio, Region is region annotation, and Reason is reason annotation of \ours{}.}
    \label{tab:system_prefer}
\end{table}

%% file: content/06_conclusion.tex
\section{Conclusion}
In this study, we analyze the region and reason for the error in the synthesized speech that makes people feel unnatural. We collect region annotation from people to gather their perception on speech unnaturalness and label regions by defined error types. Throughout the analysis, we find that the error types can cover most erroneous regions, and existing model interpretation cannot reflect these regions. We also find that the frequent type of speech unnaturalness varies between speech synthesis models, revealing model dependencies. Consequently, we show the possibility that several reasons lie under the single naturalness score, and the collection of reason annotation can show the model's weakness. 

As future work, this study can be extended to a larger dataset and application of the annotations. In order to enlarge \ours{}, we will open the dataset and annotation tool to enable the extension. For the application of the dataset, we expect a model-based evaluation that predicts human perception about the unnatural region. If the model can predict unnatural regions and reason for unnaturalness from synthesized speech, we may use the model as a new metric for speech naturalness. As shown in the preference test, we believe that using our region- and reason-annotation would enable naturalness evaluation in a more explainable way. 